\documentclass[twoside,11pt]{article}

\usepackage[T1]{fontenc}
\usepackage{apacite}
\PassOptionsToPackage{hyphens}{url}

\usepackage{jmlr2e}

\ShortHeadings{An Avalanche of Images Preceded Full-Scale Invasion of Ukraine}{Theisen, Yankoski, Hook, Verdeja, Scheirer, and Weninger }
\firstpageno{1}

\begin{document}

\title{An Avalanche of Images on Telegram Preceded Russia's Full-Scale Invasion of Ukraine}

%\author{Michael Yankoski\thanks{Colby College, \texttt{myankosk@colby.edu}} \and 
%William Theisen\thanks{University of Notre Dame, \texttt{wtheisen@nd.edu}} \and 
%Kristina Hook\thanks{Kennesaw State University, \texttt{khook2@kennesaw.edu}} \and 
%Ernesto Verdeja\thanks{University of Notre Dame, \texttt{everdeja@nd.edu}} \and 
%Walter Scheirer\thanks{University of Notre Dame, \texttt{wscheire@nd.edu}} \and 
%Tim Weninger\thanks{University of Notre Dame, \texttt{tweninger@nd.edu}} }

\author{\name William Theisen \email wtheisen@nd.edu \\
       \addr University of Notre Dame\\
       Notre Dame, IN, USA
       \AND
       \name Michael Yankoski \email myankosk@colby.edu \\
       \addr Colby College\\
       Waterville, ME, USA
       \AND
       \name Kristina Hook \email khook2@kennesaw.edu \\
       \addr Kennesaw State University \\
       Kennesaw, GA, USA
       \AND
       \name Ernesto Verdeja \email everdeja@nd.edu \\
       \addr University of Notre Dame\\
       Notre Dame, IN, USA
       \AND
       \name Walter Scheirer \email wscheire@nd.edu \\
       \addr University of Notre Dame\\
       Notre Dame, IN, USA
       \AND
       \name Tim Weninger \email tweninger@nd.edu \\
       \addr University of Notre Dame\\
       Notre Dame, IN, USA}

\editor{}

\maketitle

\begin{abstract}%   <- trailing '%' for backward compatibility of .sty file
Governments use propaganda, including through visual content—or Politically Salient Image Patterns (PSIP)—on social media, to influence and manipulate public opinion. In the present work, we collected Telegram post-history of from 989 Russian milbloggers to better understand the social and political narratives that circulated online in the months surrounding Russia's 2022 full-scale invasion of Ukraine. Overall, we found an 8,925\% increase (p<0.001) in the number of posts and a 5,352\% increase (p<0.001) in the number of images posted by these accounts in the two weeks prior to the invasion. We also observed a similar increase in the number and intensity of politically salient manipulated images that circulated on Telegram. Although this paper does not evaluate malice or coordination in these activities, we do conclude with a call for further research into the role that manipulated visual media has in the lead-up to instability events and armed conflict.
\end{abstract}


%\begin{titlepage}
%\title{An Avalanche of Images on Telegram Preceded Russia's Full-Scale Invasion of Ukraine\thanks{This work is funded by USAID under project \#7200AA18CA00059, DARPA under project HR001121C0169 and HR0011260595.}}

\section{Introduction} \label{sec:introduction}

Propaganda efforts are often used by governments to villainize and dehumanize their enemies and justify atrocities. In the age of online social media platforms these efforts are beginning to take the form of visual content that we term Politically Salient Image Patterns (PSIP) which serve to influence, demean, manipulate, and motivate various audience segments. Colloquially called memes, these image patterns---which are intended to be remixed and shared---circulate widely as a cultural reflection of world events.  However, little is known of the predictive power of PSIP. In the present work, we combine subject-area expertise on political violence and the Ukraine-Russia war with Artificial Intelligence (AI) tools from computer vision to identify Politically Salient Image Patterns (PSIPs) in visual social media that correlate with real-world political violence. This work addresses the paucity of research on how visual media may influence beliefs and opinions, an area that researchers have called a ``scientific blind spot'' \citep{weikmann2023visual}. Our analysis focuses on Russia's full-scale invasion of Ukraine in 2022, which has generated a social media narrative battle surrounding the war, from dehumanizing language aimed at justifying atrocities, competing accounts of military attacks and casualties, to heated arguments over the future of warfare itself.

Our research asks four general questions: 1) Is there an increase in image posts in temporal periods prior to the start of major political instability events? 2) If so, what is the rate of increase? These first two questions establish whether there are quantifiable numerical patterns in image posts related to a distinct political moment. We then turn to the political and technical characteristics of image patterns: 3) What kinds of \textit{Politically Salient Image Patterns} (PSIPs) emerge from our image set, and how might these PSIPs overlap with dynamics central to political instability? 4) How can we systematically identify manipulated images beyond using limited human analytic capabilities? 
From a curated list of 989 Russian \textit{milbloggers}\footnote{Milblogger refers to a person who posts online content almost exclusively about the events of an ongoing conflict, usually as part of a recruitment or propaganda strategy.} on Telegram that contributed 5,318,446 posts containing 3,264,819 images, we found an 8,925\% increase (p<0.001) in the number of posts and a 5,352\% increase (p<0.001) in the number of images in the two weeks leading up to Russia's full-scale invasion of Ukraine. Furthermore, a political analysis of these image clusters identified several threads of politically salient visual narratives that served to reinforce \textit{ingroup solidarity}, \textit{outgroup vulnerability}, and \textit{epistemic insecurity} which are all key social mechanisms that drive instability. We also found a similar increase in the number and intensity of image manipulations in that same period.

\section{Literature Review} \label{sec:literature}
Social media has facilitated democratic mobilization~\citep{steinert2017spontaneous}, but it has also become a major political tool of rumors, deception, and dehumanization~\citep{yankoski2020ai,douglas2019understanding,benkler2018network,jost2018social}. Social media is also a factor in promoting violence, and visual social media is increasingly important in political instability contexts~\citep{brown2022russia}. However, the vast majority of research on the deleterious political effects of social media has largely focused on textual analysis, not on visual media. Yang et al. \citeyear{yang2023visual} found that of 135 recent articles they surveyed on misinformation, fact-checking, propaganda, and/or fake news on social media published in seven major communication or communication-adjacent journals, only nine studies focused on images. This relative dearth of analysis is somewhat surprising given other studies which demonstrate that manipulated images have a significant influence on social media users~\citep{dan2021visual,mina2019memes,shifman2013memes,moreno2021memes,yankoski2020ai}, that \textit{DeepFake} images may upset trust and discernment in the news~\citep{patterson2019deepfake,yankoski2021meme}, and that visually-based mis/disinformation campaigns can be politically destabilizing, contributing to contested election outcomes~\citep{rosenberg2020qanon} and even bloodshed~\citep{suhartono2019violence}. The outsized impact of images appears to be because the visual aspect of a social media post carries equal or more weight in transmitting a message's intended meaning than the text included in the post\footnote{\url{https://www.darpa.mil/news-events/2019-09-03a}}. Despite the abundance of images on social media, their role and prevalence in driving instability dynamics remains understudied. 

One probable reason for the relative lack of academic analysis on visual content despite its evident importance is due to the overwhelming volume of visual social media: every minute 243,000 images are uploaded to Facebook (Aslam 2021). Every day, 500 million people post at least one image to Instagram\footnote{\url{https://investor.fb.com/financials/}}, two billion users watch or post to YouTube~\footnote{\url{https://www.youtube.com/yt/press/statistics.html}}, and hundreds of billions of photos are uploaded to social media platforms each year~\citep{eveleth2015many}. Unaided human analysis is impossible at this scale. Thus, to facilitate our analysis, we have developed sophisticated computer vision systems to assist in identifying PSIPs in visual social media at a speed and scale that would otherwise be impossible for human analysts alone. A second reason for the lack of sufficient consideration is the computational challenges images pose. The difficulty of ingesting, storing, and analyzing images—which are more computationally complex than text—is significant. And third, images pose a substantial interpretive challenge in discerning the layers of meaning, which are often context- and audience-specific, and therefore are harder to analyze at speed and at scale. Nevertheless, the enormous volume and demonstrated political importance of visual social media requires greater research attention and new partnerships across the social sciences and computer science to build systems capable of helping identify and analyze PSIPs.

In our study, the unit of analysis is the \textit{image}, whether a \textit{meme} or some other kind of image. Yet we are not interested in single isolated images, but rather in image trends: we seek to identify visual patterns that recur across different images over time. We collect, categorize, and analyze images in order to identify \textit{Politically Salient Image Patterns} (PSIPs). We attribute ``political salience'' to topics related specifically to political instability (war, mass social mobilization, etc.), and which carry significant political impact as determined by subject-area experts. Finally, we seek to determine whether images have been manipulated in some manner, without concern for whether or not the people posting or sharing the content knew that the images had been manipulated.

Ongoing technological developments in AI are helping researchers to tackle several technical challenges: \textit{scale}, where the amount of visual information on social media far exceeds the ability of human researchers to study it; \textit{forensic analysis}, or the study of image manipulation beyond what humans can perceive; \textit{provenance and distribution analysis}, examining where images originated, how they propagated across social media platforms, and are currently trending; and, eventually, \textit{semantic analysis}, models to assist human analysis of the meaning of the image itself (\textit{i.e.}, the range of possible intentions that might have motivated the creation and sharing of the image). By addressing these challenges, AI technologies can assist researchers focused on the complex dynamics of political instability. Capabilities addressing the first two challenges are more mature than those for the latter two. Thus, in this paper, we focus on the ability of AI to assist with making the scale and forensic analyses of social media images in a conflict setting more tractable. 

Focusing on the Russia-Ukraine war, we collected over 5.3M posts and 3.2M images posted between October 2015 and March 2023 from 989 Telegram accounts whose regular content focuses heavily on Russia/Ukraine relations. Telegram is one of the most popular social media platforms in this context. Section V of the present work details our research design and methods, but in short, we conducted unsupervised clustering (\textit{i.e.}, clustering based on semantic image similarities, without human labeling or intervention) on a subset of images posted to Telegram channels between February 10, 2022 and March 10, 2022, \textit{i.e.}, the month immediately surrounding Russia's full-scale invasion of Ukraine on February 24, 2022, allowing us to focus on a smaller temporal window where patterns could be more effectively studied. The clustering process surfaced groups of similar images that occurred more frequently than others. A subset of these image clusters was then analyzed by our team's subject-area experts to identify PSIPs. Our team includes computer scientists specializing in AI as well as social scientists with expertise on Ukraine/Russia dynamics. 

Identifying PSIPs in such a large set of images requires a combination of AI systems to cluster images and subject-area experts to interpret the meaning of the images. Subject-area experts can identify and interpret image narratives by drawing on deep contextual knowledge of political and social conditions. However, AI's ability to organize and cluster enormous amounts of social media data in near real-time enhances expert identification of trends that may be otherwise missed by conventional, smaller-scale analyses by humans, or suffer from selection bias because of their limited scale. Our approach overcomes some of the limitations and potential biases of other approaches by increasing the amount of data that is analyzed. This is important not only quantitatively in that there is more information available, but qualitatively in that it introduces a large body of different kinds of data, \textit{i.e.}, visual information. This in turn may give researchers additional insights into unfolding dynamics as well as more confidence in their findings.

\section{The Russia-Ukraine War} \label{sec:war}

The Russia-Ukraine armed conflict began in 2014, with Russia's seizure of Ukraine's Crimean Peninsula and armed conflict through proxy forces in eastern Ukraine. These events occurred only days after state-initiated killings had brought months of protests in Ukraine's Revolution of Dignity (Euromaidan) to a climax~\citep{shore2018ukrainian}. Ukrainian President Viktor Yanukovych's implication in the killings prompted him to lose his parliamentary majority, and he fled to Russia~\citep{marples2014ukraine,mcdonald2016ukraine}. Ukraine's parliament passed bills restoring the country's 2004 constitution, formally removing Yanukovych from office, calling for snap elections in three months, and declaring their intentions to embark on a domestic reform agenda committed to European integration\citep{fisher2014ukraine,marples2014ukraine,mcdonald2016ukraine}. 

Russia captured strategic military locations in Ukraine by February 27, 2014~\citep{yekelchyk2015conflict} and quickly installed a puppet government, staged a falsified referendum, and fueled conflict in southern and eastern Ukraine, using regular Russian troops without identifying insignia, the disbanded Ukrainian Berkyt riot police, and some local volunteers, often linked to organized crime~\citep{galeotti2016hybrid,dunn2014empire,grau2015brothers}. The first eight years of the Russia-Ukraine armed conflict ushered in enormous political, economic, environmental, and human consequences for Ukraine. The United Nations estimates that between 2014-2019, more than 13,000 lives were lost, 30,000 were wounded, and 1.5 million people internally displaced~\footnote{\url{https://www.unian.info/war/10416549-donbas-war-death-toll-rises-up-to-nearly-13-000-un.html}}.

Russia's full-scale invasion of Ukraine (referred to by Russian propagandists as a ``special military operation'') began on February 24, 2022~\citep{garmone2023ukraine}. While our case study is centered on the two weeks before and after this date, eight years of armed conflict prior to the full-scale invasion provide essential context. Given years of armed conflict and numerous documented war crimes in Russian-controlled areas from 2014-2021, a sense of foreboding is present in the media context in this timeframe, without the full knowledge that hindsight provides today.

\section{Hypotheses and Research Design} \label{sec:design}

We define an \textit{image} as a single visual artifact, whether manipulated or not, posted by a user, and circulating on social media. This image may be a meme, photograph, illustration, video screenshot, other visual representation, or a combination of these. It is the smallest unit of analysis. A meme is an image that is designed to be shared and remixed by others~\citep{mina2019memes,theisen2021automatic,shifman2013memes}. 

Our research efforts are focused on identifying politically salient image patterns in contexts of political instability.  \textit{Political instability} refers to single or combined events that represent a significant shock or threat of shock to the political system. We focus on contexts marked by governance or security crises, as they are among the most severe types of political instability shocks~\citep{goldstone2010global}. These include the onset or intensification of armed conflict (civil war or international war), direct threats to leadership integrity (successful or attempted assassinations and coups), mass atrocities including but not exclusively killings, and/or widespread and sustained civilian mobilization against the state. The quantitative conflict research literature provides a wide range of definitions of instability and related phenomena using generally high numerical thresholds of mass killings or similar indicators, and often large temporal windows for analysis (month or year). Those datasets are often temporally lagged, which provide little analytical purchase for understanding contemporary or near-contemporary instability trends~\citep{goldstone2010global,harff2017no}. Although existing instability conceptualizations are useful for analyzing general long-term trends, we are interested in complex, short-term and highly fluid instability dynamics. Thus, we identify political instability events by drawing on subject-area experts from our own team and interviews with other country and subject-area experts.  Importantly, a political instability event is defined separately from whether it appears on social media, to avoid endogeneity issues.

We generated several testable hypotheses for this project. 

\textit{H1: There will be an increase in posts containing images on selected Russia/Ukraine-relevant Telegram accounts prior to and after major political instability events.}

Prior research suggests that misinformation and propaganda spreads rapidly on social media during times of major political instability~\citep{zeitzoff2017social}. Given the political volatility of relations between Ukraine and Russia at this time, we anticipate a rise in social media posts containing images on social media corresponding to the onset of Russia's full-scale invasion. While it may seem self-evident that posts containing images would increase around the escalation of violence, it is important to establish and quantify this increase.

\textit{H2: The rate of increase in posts of images on selected Telegram accounts will be higher closer to the onset of major political instability events. }

As crises escalate, social media becomes an increasingly important domain for information warfare and efforts to shape narratives and interpretations of ongoing events~\citep{zeitzoff2011using}. These dynamics suggest we will see not only a rise in the total number of images posted, but also a rise in the rate of images posted on politically prominent Telegram accounts. 

Positive confirmation of H1 and H2 will provide a foundation for the more substantial analyses proposed in H3 and H4.

\textit{H3: Clusters of images will evince ingroup solidarity (IS), outgroup vulnerability (OV), and epistemic insecurity (EI), which are central to political instability dynamics.}

Although social media posts do not cause instability, research suggests social media impacts instability dynamics through three broad mechanisms. The first is by reinforcing \textit{outgroup vulnerability} (OV), which can include spreading dehumanizing or polarizing discourse about opponents and framing them as existential threats~\citep{asmolov2018disconnective,haigh2019information}. The second is by reinforcing \textit{ingroup solidarity} (IS) and cohesion by positioning one's group as defenders of core values or survival, or by framing a collective identity through a sense of shared grievances, historical wrongs, or fear~\citep{rio2021myanmar}. The third is through the spread of \textit{epistemic insecurity} (EI), the cumulative effect of sustained charges of bias or conspiracy that erode the status of truth claims, standards of evaluation, and evidence, in favor of unsubstantiated narratives~\citep{boyte2017analysis,yankoski2021meme}.

The efficacy of social media in spreading political information and shaping ingroup and outgroup identities leads actors to focus their efforts on the range of topics that are of high political salience~\citep{allington2020conspiracy}. Therefore, we anticipate finding image patterns that attempt to influence instability dynamics through OV, IS, and EI.

\textit{H4: A statistically significant number of the images from the days immediately preceding and following a major political instability event will have more manipulated pixels compared to the same time period the year before.}

The potential effect of manipulated imagery on social media users may be willfully harnessed by leaders and used to strategically influence popular opinion in the lead-up or aftermath of a political event. Small manipulations can have enormous implications in conflict settings: an image manipulated to suggest the assassination or imprisonment of an important political figure, the dehumanization of groups to lower resistance to violence against them, the surrender of a key city or geographic region, and/or a new ceasefire could all have enormous impacts. Thus, in heightened crisis moments we expect to see higher rates of manipulation, which would suggest more intense struggles over shaping popular narratives and understandings of events. These efforts could take a variety of forms, such as large image manipulations to highlight an important political claim (including satire/caricature), memes using context-specific wordplay, and references resonant with one community of insiders but whose meaning may be lost on others.

As seasonal patterns that influence the rates and types of social media posts have been noted by researchers~\citep{villamediana2019destination}, we will compare the number of manipulated images posted during the start of the full-scale invasion (February 2022) with the number of manipulated images posted a year prior (February 2021) and a year after (February 2023).

\section{Methodology} \label{sec:methodology}

Evaluating \textit{H1} and \textit{H2} requires a numerical analysis of our dataset on the basis of the date on which the images were posted to Telegram. We collected the entire history of all posts and images posted by the Telegram accounts of 989 Russian milbloggers, which were identified by subject matter experts as focusing on Russia/Ukraine relations. Our dataset as of May 23, 2023 consists of 5,318,446 Telegram posts and contains 3,264,819 images. The posts range in date from mid-2015 to May 2023; collection of posts from these accounts is ongoing. From this larger image set, clustering was performed on the subset of 144,048 images posted between February 15, 2022 and March 15, 2022, a date range immediately preceding and following Russia's full-scale invasion of Ukraine, which began on February 24, 2022. The null hypothesis is that the raw number of posts containing images and the posting rate will remain statistically unchanged from before the full-scale invasion compared to after the full-scale invasion.

Evaluating \textit{H3} required the identification and analysis of image \textit{clusters} — groups of images that have some visual similarity — which appeared on Telegram during the specified time period. We utilized the image clustering pipeline method outlined in Theisen et al. \citeyear{theisen2023motif} to cluster the aforementioned subset of 144,048 images. Our clustering pipeline functioned as follows: first, image features were extracted from each individual image using the MobileNet algorithm~\citep{howard2019searching}. These image features were then stored using Facebook's AI Similarity Search (FAISS) engine~\citep{johnson2019billion}. Using FAISS as a query engine, image feature similarities were computed and similarity scores between images were used to generate a similarity graph of the images. Clusters of images were then identified from the graph of images using spectral clustering~\citep{shi2000normalized}. This clustering process produced 1,988 clusters with an average of 72 images per cluster.
In addition to the algorithmic clustering and numerical analysis of the images, evaluating H3 also required subject-area expert analysis of a random subset of the clusters to evaluate for political salience. 362 of these clusters (approximately 18\%) were randomly selected and further analyzed by two Russia / Ukraine conflict experts with Ukrainian and Russian language fluency. The analysts separately performed a two-stage analysis of this subset of image clusters to:

\begin{enumerate}
    \item Identify a center of the cluster, \textit{i.e.}, the most numerically significant and discernibly similar group of images within that specific cluster. While reviewing a cluster of dozens or hundreds of images, the center of that cluster was operationalized as the sub-group of images that were the most evidently similar to one another and which occurred the most frequently within the cluster. Often this center was a single image that had been re-posted by multiple Telegram accounts, and thus appeared multiple times within a single cluster.
    \item Identify whether the center of a particular cluster was politically salient. The concept of political salience was operationalized after considering two distinct potential interpretations: a) as defined by conflict-affected victims (\textit{e.g.}, a meme celebrating the downing of an individual plane might be considered politically salient by many Ukrainians, given the credible fears of the aircraft targeting civilians and cities) and b) as defined by an external observer of the full-scale invasion, for whom politically salient referred to significant events that might influence the war's overall trajectory. Images which were coded as politically salient needed to rise above events that might be reported daily in an active war zone (\textit{e.g.}, reporting of an air raid alarm, unless this occurrence was unusual in some way). This allowed multiple interpretations from various perspectives, while keeping the category focused enough to retain conceptual meaning.
\end{enumerate}

Evaluating \textit{H4} required the utilization of image manipulation detection algorithms over the entire dataset of images collected from Telegram. Two algorithms were used to test every image in the dataset for manipulations: Content Aware JPEG Inconsistencies (CAGI-INV) and blind local NOIse estimation (NOI2)~\citep{vidalmata2023effectiveness}. All 144,048 images from the date range used in the above clustering experiments (February 15, 2022 to March 15, 2022) were processed by both 
algorithms, and the output heatmap of each image was converted to a binary (black and white) image and the ratio of the number of edited to non-edited pixels was calculated. The null hypothesis is that the number of edited images will remain unchanged from before the full-scale invasion compared to after the full-scale invasion.

\section{Findings}

\subsection{H1 and H2: Numerical Analysis of Posts}

We found that there was a significant increase in the total number of Telegram posts (both text-based posts as well as posts containing images) by Russian milbloggers leading up to the full-scale invasion. To answer this question in statistical terms, we performed an interrupted time series (ITS) quasi-experiment using an autoregressive integrated moving average (ARIMA) to control for temporal effects. This analysis compared the number of posts and images posted by Russian milbloggers before and after the full-scale invasion.

\begin{figure}[t]
    \centering
    \includegraphics[width=.95\textwidth]{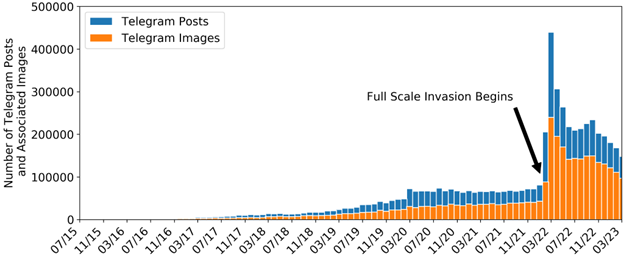}
    \caption{Raw number of collected Telegram posts and their associated images from 2015 to 2023.}
    \label{fig:fig1}
\end{figure}

Figure~\ref{fig:fig1} illustrates an increase in the number of posts, both text-based posts as well as posts containing images, in the weeks and months surrounding Russia's full-scale invasion. Overall, we find that the number of posts and images steadily increased at a rate of 1.36\% and 0.76\% (p<0.001) per day respectively before Russia's full-scale invasion. Comparing February 15, 2022 and the start of Russia's full scale invasion on February 24th, 2022, we observed an increase of 8,925\% (p<0.001) in the number of posts and an increase of 5,352\% (p<0.001) in the number of images. Interestingly, the total number of posts and images has steadily decreased over the time frame studied by 19.69\% and 9.40\% (p<0.001) respectively per day after the full-scale invasion. Thus, H1 is confirmed, as we see a statistically significant increase in both the number of text-based posts as well as image-posts by Russian milbloggers associated with Russia's full-scale invasion.

We also confirmed that the rate of increase of image posts would rise closer to the onset of a major instability event, confirming H2. 

\begin{figure}[t]
    \centering
    \includegraphics[width=.95\textwidth]{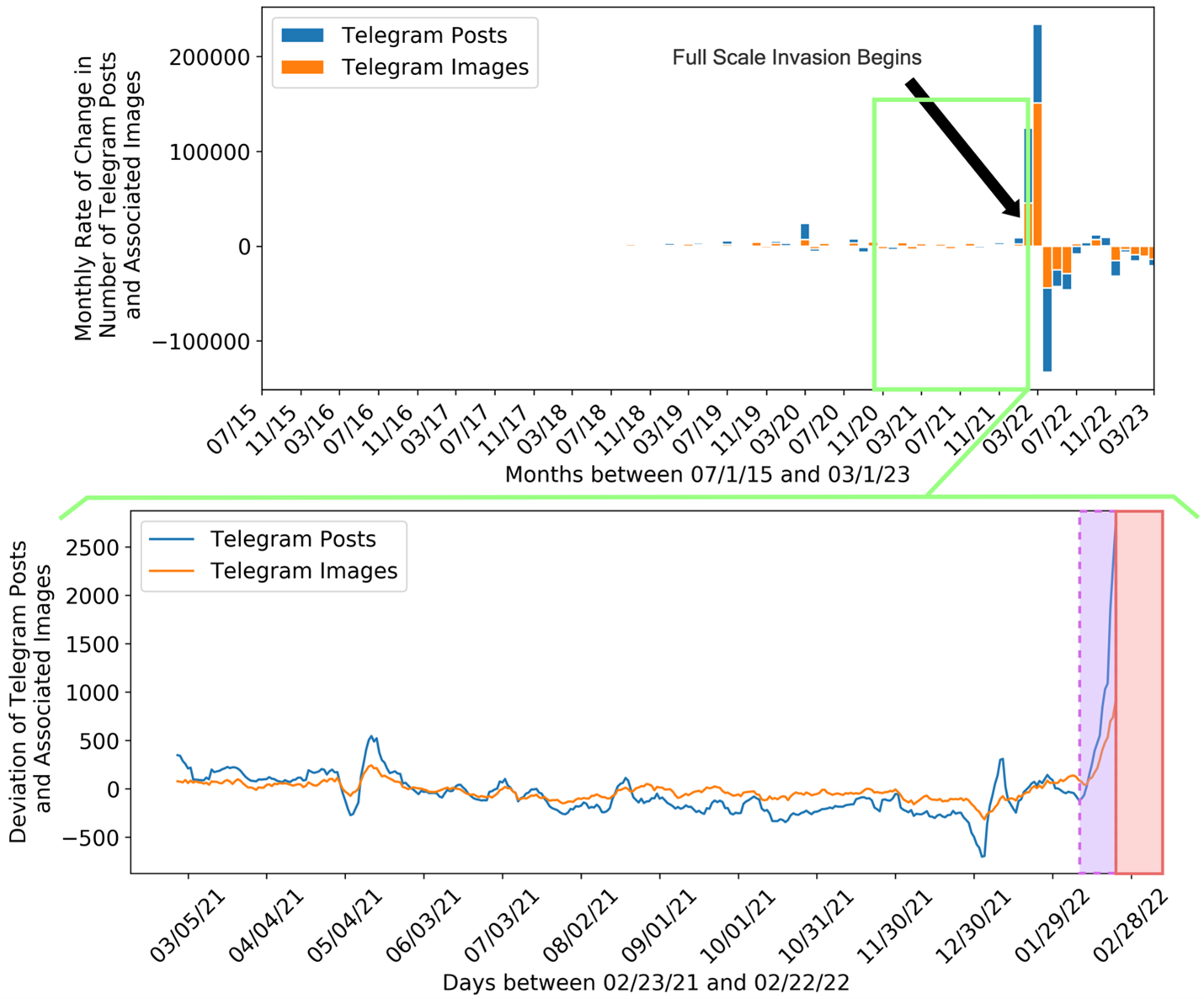}
    \caption{Month-to-month difference in total number of posts and posts containing images. A spike in activity corresponds with the beginning of Russia's full-scale invasion in the Ukraine.}
    \label{fig:fig2}
\end{figure}

Figure~\ref{fig:fig2} illustrates the daily deviation from the trendline computed from the OLS analysis before the full-scale invasion began. From the inset illustration in Figure 2, we note a stark departure from the trendlines. To investigate this further, we performed changepoint analysis comparing the deviation from the trendline for various time-windows preceding Russia's full-scale invasion. We detected an inflection point showing that Russian milbloggers dramatically increased the rate of posts and image-posts on February 4th, 2022, nearly three weeks prior to Russia's full-scale invasion. Between February 4th and 22nd, posts increased 259\% per day and image-posts increased at a rate of 76\% per day (each day on average), a significant rise.

\subsection{H3: Analysis of Image Clusters}

The subset of 362 image clusters, randomly selected, out of the total of 1,988 were analyzed by the subject-area experts separately before conferring about results, to help ensure intercoder reliability~\footnote{Intercoder reliability was 98\%: 6 out of 362 clusters were categorized differently. The analysts discussed these remaining six clusters to evaluate the political salience and resolve differences. For example, regarding one cluster concerning a Ukrainian downing of a Russian aircraft, one analyst coded the cluster as ``yes'' (politically salient) while the other analyst coded it as ``no,''meaning it did not rise to the level of being unusual in a daily wartime context, as several hundred Russian aircraft have been shot down.}. 

Ultimately, subject-area experts identified 33 clusters as having a politically salient center. Although analyzing the specific social impact of these politically salient clusters is beyond the scope of our project, our experts coded these clusters as operating through three mechanisms (presented in H3): ingroup solidarity (IS), outgroup vulnerability (OV), and epistemic insecurity (EI). Importantly, any given image within a cluster may operate through more than one mechanism, \textit{e.g.}, an image may promote IS and EI simultaneously.

\begin{figure}[t]
    \centering
    \includegraphics[width=.75\textwidth]{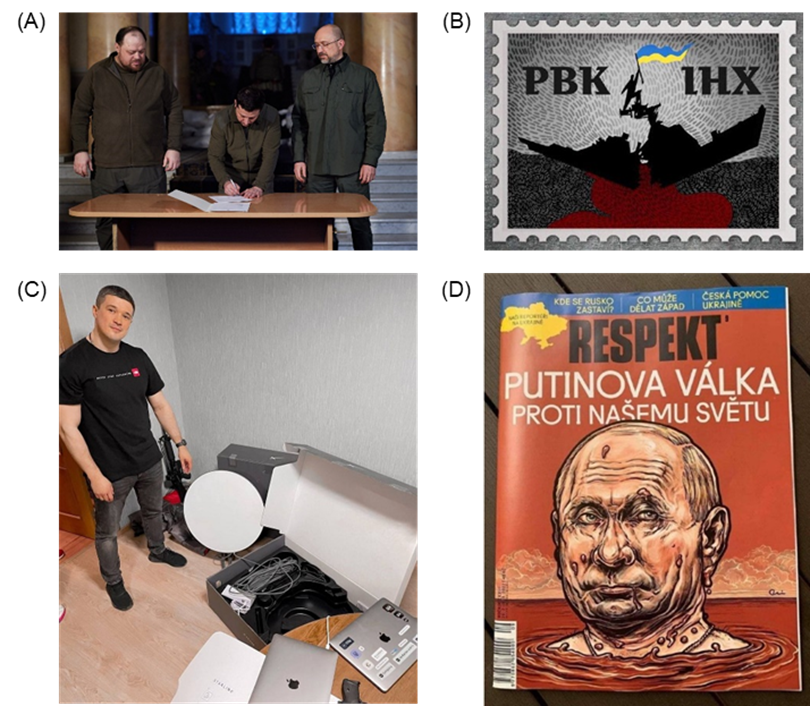}
    \caption{Politically salient image exemplars at the time of the Russian full-scale invasion of Ukraine.}
    \label{fig:fig3}
\end{figure}

Here we present a subset of the politically salient images emerging at the center of their respective clusters. Figure~\ref{fig:fig3}A shows Ukrainian President Volodymyr Zelenskyy, flanked by Prime Minister Denys Shmyhal, and Verkhovna Rada (Parliament) Chairman, Ruslan Stefanchuk, affirming Ukraine's decision to pursue NATO membership. This image functions along both IS and OV mechanisms. Figure~\ref{fig:fig3}B is a stamp referencing the Russian warship Moscow, which ordered a small Ukrainian detachment to surrender or die; the Ukrainian service responded, ``Russian warship, go fuck yourself,'' which has become an internationally recognized slogan in the war. This image operates along mechanisms of IS and OV. Figure~\ref{fig:fig3}C is a photograph of Ukrainian Minister of Digital Transformation Mykhailo Federov posing with Starlink Internet technology, reinforcing IS and OV. Figure 3.D is the cover of the Czech magazine, Respekt, widely circulated in Ukraine, and depicts Russian president Vladimir Putin as up to his neck in blood, in reference to Russia's civilian targeting and war crimes. It states, ``Putin's War against Our World'' in Czech, and functions along the IS and OV mechanisms.\footnote{\url{https://war.obozrevatel.com/voyuet-protiv-vsego-mira-cheshskoe-izdanie-respekt-razmestilo-na-oblozhke-putina-v-krovi-foto.htm}}

\begin{figure}[t]
    \centering
    \includegraphics[width=.75\textwidth]{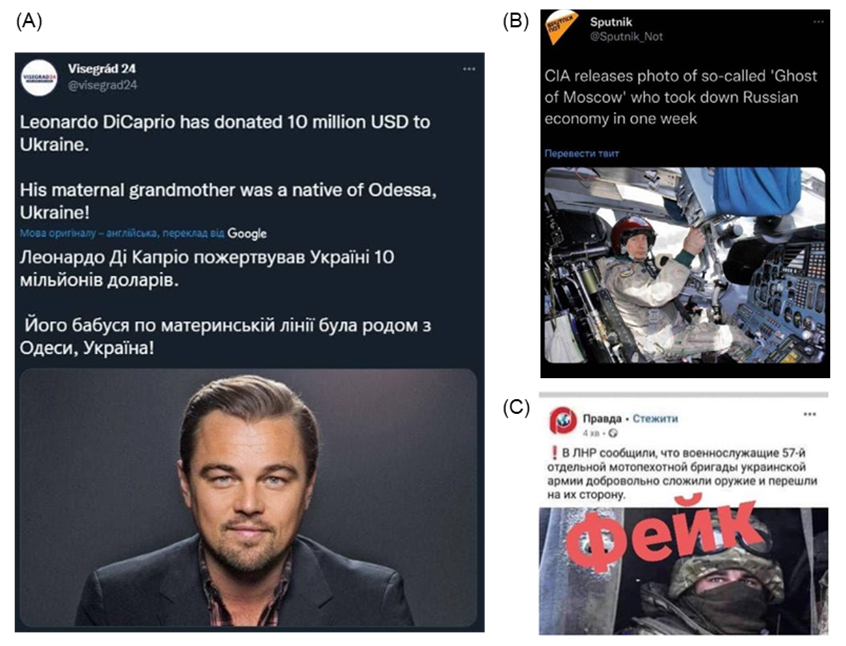}
    \caption{Politically salient images at the time of Russia's full-scale invasion of Ukraine that are fake or misleadingly out of context.}
    \label{fig:fig4}
\end{figure}

Figure~\ref{fig:fig4} presents several politically salient images found at the center of their respective clusters, which are fake or present something out of context. In addition to promoting IS and OV, these images also spread epistemic insecurity through false or misleading representations.  Figure~\ref{fig:fig4}A depicts a false rumor that actor Leonardo DiCaprio had donated \$10 million dollars directly to Ukraine, given a familial tie there. Both the donation as well as the story about his family connection to Ukraine are false, but this rumor spread ``across the world''\citep{dale2022fact}. Figure ~\ref{fig:fig4}B references a rumor spread in the early weeks of the war about a fighter pilot dubbed the ``Ghost of Kyiv'' who was credited with protecting the skies during Russia's Kyiv offensive, shooting down multiple Russian planes. This meme---with Russia's president Putin's face edited into the image as the pilot---pokes fun at the spectacular early failures of the Russian military and Russia's growing isolation, naming Vladimir Putin as bearing responsibility for declines in the Russian economy. Figure~\ref{fig:fig4}C shows a fake message stating that the members of the well-known Ukrainian 57th Brigade surrendered to the Russians. Other versions of this rumor stated terrible losses or territorial encirclements of the brigade, causing a stir in Ukrainian media networks before the commander publicly refuted the narrative and fake images. 
The above image examples function in at least three politically salient ways: reinforcing IS and OV while also promoting EI. These images also demonstrate the power of visual media — and in particular the potential reach of PSIP— as four different languages (Ukrainian, Russian, English, and Czech) are present in these visual artifacts even as their content allows their core messages to be interpreted without translation. 

\subsection{H4: Analysis of Manipulated Images}

The difficulty of interpreting manipulated images is evident in an earlier period of the conflict when Russia falsely denied any military presence or support for proxy forces~\citep{dunn2014empire,yekelchyk2015conflict}. A sustained information war involved the use of electronic communication channels to organize participants and sympathizers on both sides~\citep{economist2015battle}. 

\begin{figure}[t]
    \centering
    \includegraphics[width=.75\textwidth]{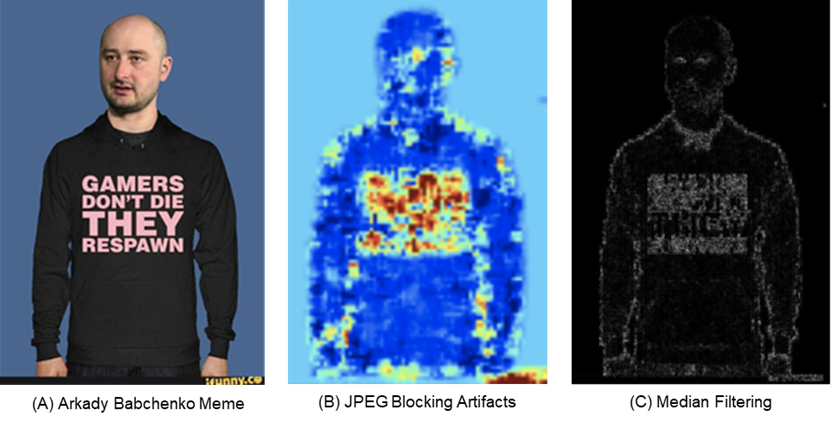}
    \caption{(A) Pro-Russian meme mocking anti-Putin journalist Arkady Babchenko. The text on the shirt was inserted into the photo. Babchenko's head appears to be inserted onto someone else's body. (B) JPEG blocking artifacts from alterations of the image; (C) Median Filtering results reveal alterations of the image. Media forensics tools can detect the tampering~\citep{lin2009fast}, but cannot understand its political relevance.}
    \label{fig:fig5}
\end{figure}

Figure~\ref{fig:fig5} shows a pro-Russian meme mocking anti-Putin journalist and former Russian soldier Arkady Babchenko, whose death was faked by Ukrainian security services in order to expose an assassination plot against his life~\citep{macfarquahar2018after}. This meme was downloaded from iFunny, an online imageboard hosting numerous memes associated with militant and extremist movements\citep{broderick2019ifunny}. The language ``gamers don't die they respawn'' is an allusion to the programmed behavior of video game characters who spontaneously return to life to continue gameplay, making light of Babchenko's predicament. While manipulation detection algorithms can detect the manipulations that have been performed on the image, AI is not yet capable of understanding the image's multiple layers of relevance in complex political situations like war and instability.

Although interpreting multifaceted meanings of image manipulations lies beyond the current state of the field, our AI image processing pipeline system can identify images that have been manipulated in some way. Our system can do this at scale and a level of technical precision beyond what human analysts can accomplish. To determine the extent to which images in the dataset have been manipulated an experiment was run. All 144,048 images between 02/15/22 and 03/15/22 were processed by the two manipulation detection algorithms. Rather than manually reviewing images, the algorithm counted the number of images with a percentage of pixels edited to the nearest ten percent.

\begin{figure}[t]
    \centering
    \includegraphics[width=.95\textwidth]{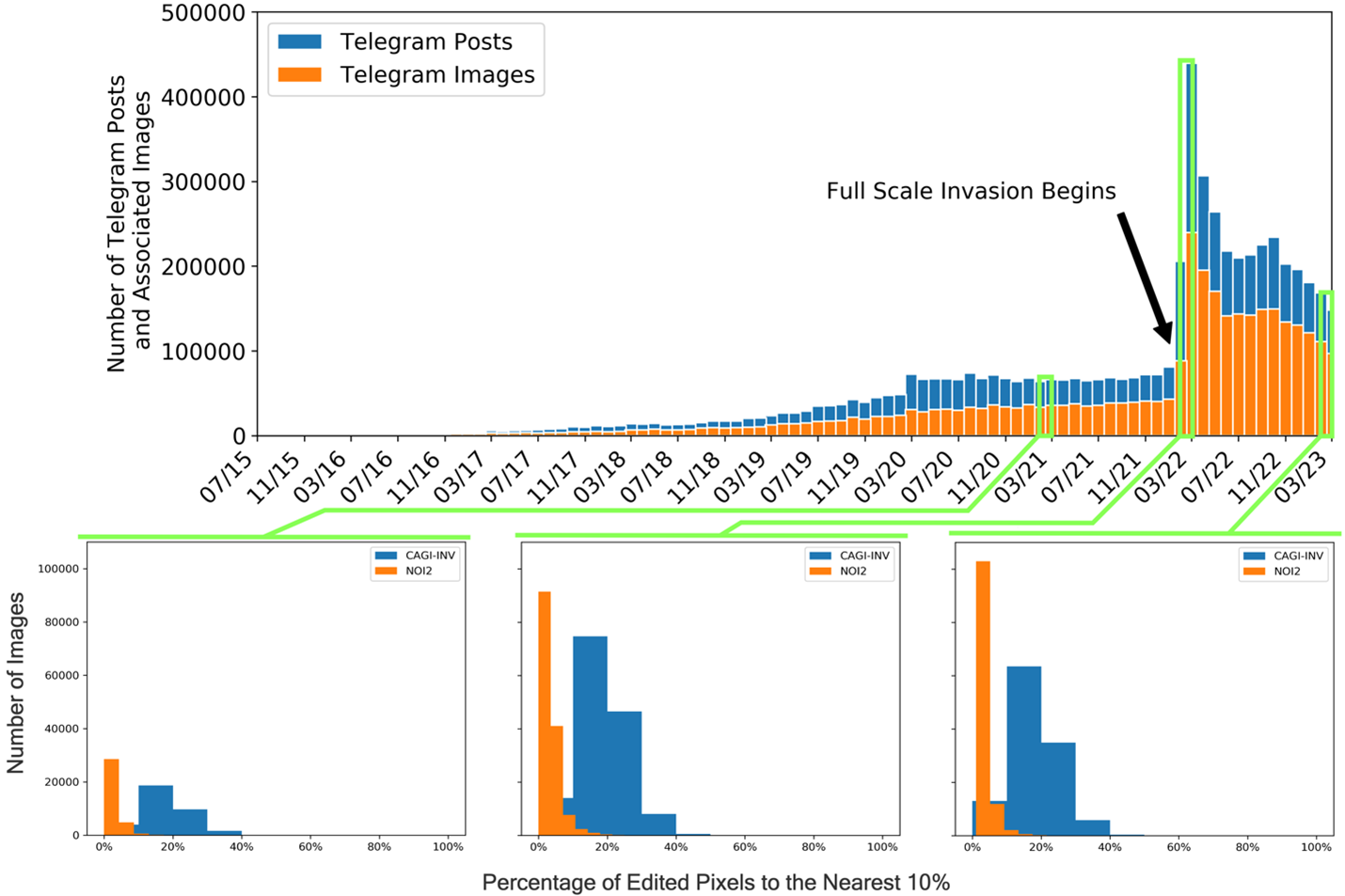}
    \caption{(Top) Number of posts collected from Telegram shifts dramatically during Russia's full-scale invasion of Ukraine. (Bottom) The number of edited images also increases dramatically. The distribution of percentage of edited pixels images detected by two different media forensics tools (CAGI-INV and NOI2) for one-month spans: (left) one year prior to Russia's full-scale invasion of Ukraine (02/15/21 - 03/15/21), (middle) during the full-scale invasion (02/15/22 - 03/15/22), and (right) one year after the full-scale invasion (02/15/23 - 03/15/23). Analysis shows that the number of alterations is significantly higher during and after the full-scale invasion compared to the year prior (HSD=0.709, p<0.001).}
    \label{fig:fig6}
\end{figure}

A distribution of how many images fell within each ten percent range can be seen in Figure~\ref{fig:fig6}(Bottom). Note that there is no a priori threshold for how many of an image's pixels need to be edited for that image to actually have been manipulated, and the percentage of edited pixels for the examples below is much smaller than what might be expected to dramatically alter the range of potential meanings of the image. 

In addition to all 144,048 images used in the experiments above, we tested all images in the same range (02/15/2022 - 03/15/2022) from a year prior to the full-scale invasion (34,105 images) and all images in the range a year after the full-scale invasion (117,714 images), to provide temporal context for the numbers and distribution of edited images. Building upon our findings in H1 that the month surrounding the full-scale invasion has the largest number of images of the three time periods, we found the following regarding H4: not only were more images posted during this volatile period, but more of the pixels in the images posted during this time period were edited. An ANalysis Of VAriance (ANOVA) test found the distributions were significantly different for CAGI-INV (F(3,295867)=272.58, p<0.001) and NOI2 (F(3,295867) =1066.95, p<0.001). The pairwise Tukey HSD test showed that the one-month time period surrounding the full-scale invasion had a statistically higher percentage of edited pixels per image compared to the same one-month interval the year prior (HSD=0.709, p<0.001) and compared to the same one-month interval the following year (HSD=0.557, p<0.001). We therefore reject the null hypothesis and conclude that there is a higher rate of image alterations surrounding the full-scale invasion.

\subsection{Example Manipulations}

\begin{figure}[t!]
    \centering
    \includegraphics[width=.95\textwidth]{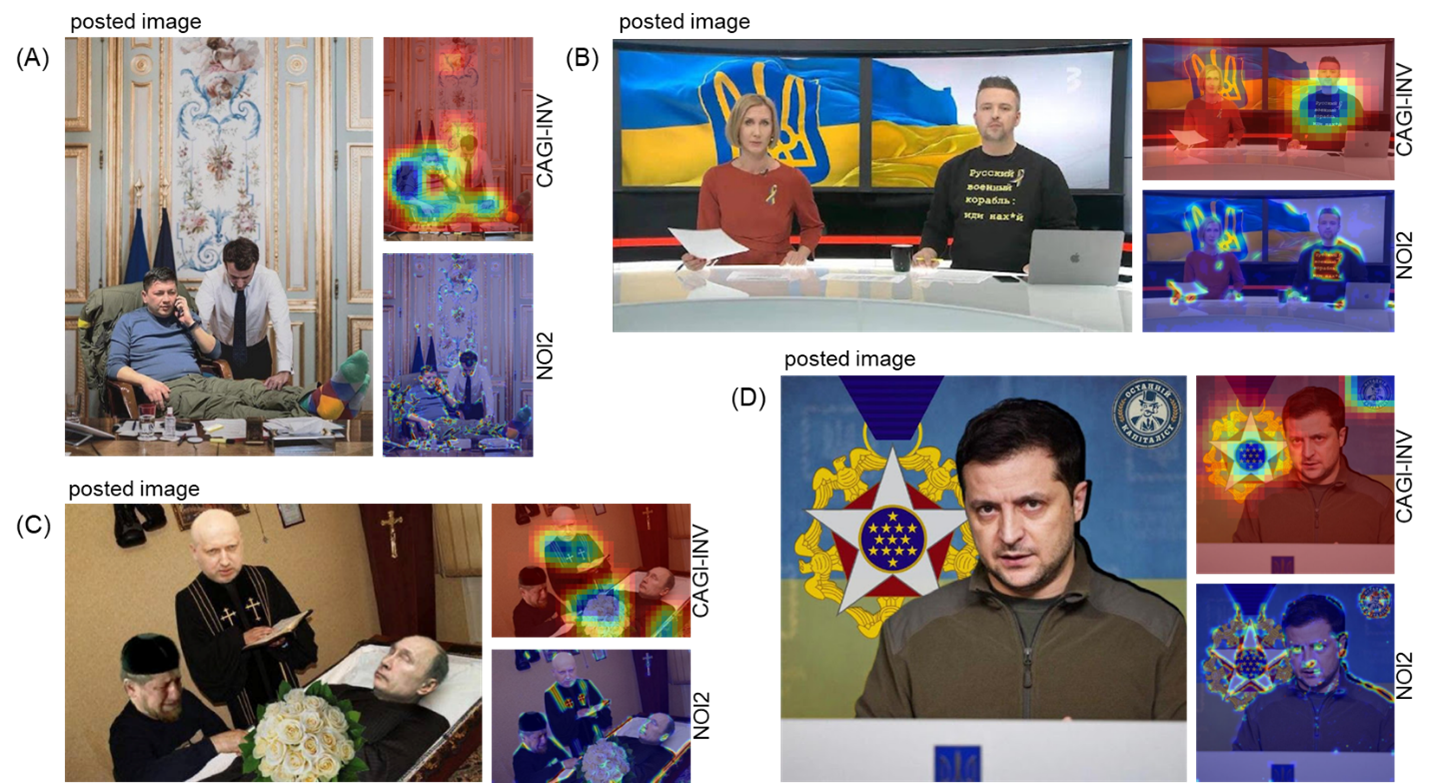}
    \caption{Example images widely spread on social media that were altered to convey a political message. (A) French president Macron (in context signaling defeatist attitude of Western Europe or appeasement) composed with a photo of Mykolaiv governor Vitality Kim in socks (known for his funny threats to Russians). (B) This picture purportedly depicts the anchor of a Latvian television show wearing a shirt with a famous wartime resistance slogan telling off a Russian warship. While this slogan appears legitimately in many types of products and clothing, this instance suggests image manipulation. (C) Putin falsely depicted as deceased, while being mourned by Ramzan Kadyrov, leader of the Chechen Republic. Presiding over the funeral as the priest is Ukrainian politician Oleksandr Turchynov, who served as Acting President of Ukraine from February to June 2014. (D) This image features Ukrainian president Volodymyr Zelenskyy in front of a U.S. Presidential Medal of Freedom, the highest civilian award granted by the United States. As of this writing, Zelenskyy has not received this award.}
    \label{fig:fig7}
\end{figure}

Figure~\ref{fig:fig7} shows examples of manipulated images and the heat-map of manipulated pixels identified by our detection system. We include these examples from our complete set of images that contain manipulated pixels to illustrate the potential political significance of manipulated images. Among the manipulations we detected in our image set, the images included here are some of those that are particularly noteworthy due to their political salience. The image with 
Macron suggests Ukrainians' outsized courage amid foreign leaders who appeared afraid of Russian nuclear threats. Additionally, the meme of the Latvian newscaster portrays a scenario with a slogan commonly appearing on clothing, yet manipulated in this specific instance. The image of Ukrainian president Volodymyr Zelenskyy has a backdrop of the U.S. Presidential Medal of Freedom, the highest civilian award granted by the U.S. Zelenskyy has not received this award at the time of this writing. As noted above regarding multiple possible interpretations of images, this image may be suggesting Zelenskyy's symbolic role in defending freedom that was especially prominent during February 2022, or it may be mocking Zelenskyy as a puppet of US influence in the region. The image of the deceased Putin is especially noteworthy. In this image Putin is pictured as deceased, while being mourned by kneeling loyal subordinate Ramzan Kadyrov, leader of the Chechen Republic. The priest is Ukrainian politician Oleksandr Turchynov, former Acting President of Ukraine from February to June 2014 following President Viktor Yanukovych's flight to Russia. Turchynov is an Evangelical Protestant (a religious minority of less than 2\% in Ukraine), who serves as an elder and lay minister in his Baptist church. With the Evangelical church targeted for persecution and suppression within Russia, some Russian-affiliated propaganda targeted him with the insulting nickname, ``Bloody Pastor.'' However, this nickname was viewed by Turchynov and broader Ukrainian society as a source of irony and laughter, skyrocketing his popularity in Internet memes. This meme was likely created during one of Putin's temporary absences from public life, expressing the hope that he perished and joking that Turchynov would step in again as caretaker leader, this time of Russia.

\section{Conclusion} \label{sec:conclusion}

Politically salient image patterns (PSIPs) on social media are an important lens through which to analyze real-world conflict and instability dynamics. Our findings are significant, including a sharp rise in visual posts prior to a major instability event and an increase in the \textit{rate} of postings closer to instability event onset. We also find that utilizing AI image clustering techniques to help parse vast image sets makes it possible for experts to identify PSIPs that promote ingroup solidarity, outgroup vulnerability, and epistemic insecurity. Lastly, we find that image manipulation patterns increase in the periods around major instability events, suggesting the importance political actors give to shaping and controlling narratives that circulate on social media. What we have presented is not simply a call for the technical extension of current social scientific work on social media, nor merely the contribution of more data to already existing analytical models of conflict and instability. Rather, we suggest that the systematic analysis of images posted to social media can make several direct contributions to peace and conflict research.

First, focusing on images not only increases the amount of data available to researchers but it also adds a new class of data that has not been explored systematically at scale. Although text is much easier to collect and analyze, different forms of social media have variable political effects given their comparative sophistication and subtlety. For example, it is much more effective to spread manipulated but highly realistic images of political opponents engaged in violence than it is to make substantiated claims with text~\citep{woolley2020reality}.

Second, new data on images can also generate novel hypotheses about how the ideological framing of adversaries and campaigns to undermine truth can shape instability onset, escalation, and spread, as well as hypotheses about the contexts under which such influence efforts may fail. Exploring how images on social media exacerbate instability requires further analysis. The need for advanced and real-time multi-modal systems that are capable of organizing and interpreting images in order to identify politically salient social media campaigns is likely to increase, especially those dealing with more sophisticated and subtle forms of manipulation and influence. The collaborative image analysis systems our team has built represent a technical advance in the field of computer science and computer vision and promising collaborations. The challenge of interpreting at larger scales the subtle meanings and potent influences of PSIPs requires significant additional research, a task underscored by the rate of data production in real-time. Our paper suggests one path forward for doing so.  

\section*{Acknowledgements}
This work is funded by USAID under project \#7200AA18CA00059 and DARPA under projects HR001121C0169 and HR0011260595.

%\bibliographystyle{apa}
%\bibliography{bibliograph}

\end{document}